\documentclass[journal]{IEEEtran}

\usepackage{amsmath,amssymb,amstext} % Lots of math symbols and environments
\usepackage{bbold}  % for identity matrix \mathbb{1}
\usepackage[pdftex]{graphicx}
\usepackage{caption}
\usepackage{subcaption}
\usepackage{multirow}
\usepackage{array}
\usepackage{float}  % for [H]
\usepackage{url}
\usepackage{xspace}  % for space after macro
\usepackage{enumitem}  % custom enumerate labels
\usepackage{authblk}  % author block
\usepackage{rotating}
\usepackage{pdflscape}
\usepackage[disable]{todonotes}  % [disable] to suppress
\usepackage[final]{changes}  % \added{}, \deleted{}, \replaced{good}{bad}. Use option [final] to clear traces.

\newcommand{\mat}[1]{\mathbf{#1}}
\renewcommand{\vec}[1]{\mathbf{#1}}
\DeclareMathOperator*{\argmin}{arg\,min}

\newcommand{\petco}{P$_{\text{ET}}$CO$_2$}

\title{Monocular 3D Sway Tracking for Assessing Postural Instability in Cerebral Hypoperfusion During Quiet Standing}

\author{Robert Amelard,~\IEEEmembership{Member,~IEEE,}
        Kevin R Murray,
        Eric T Hedge,
        Taylor W Cleworth,
        Mamiko Noguchi,
        Andrew C Laing,
        Richard L Hughson% <-this % stops a space
\thanks{This work was supported by the Natural Sciences and Engineering Research Council of Canada (PDF-503038-2017). (\textit{Corresponding author: Robert Amelard})}
\thanks{R. Amelard is with the Schlegel-UW Research Institute for Aging, Waterloo ON N2J 0E2, Canada (e-mail: ramelard@uwaterloo.ca)}% <-this % stops a space
\thanks{KR Murrary, ET Hedge, A Laing, and RL Hughson are with the Schlegel-UW Research Institute for Aging, Waterloo ON N2J 0E2, Canada and Department of Kinesiology, University of Waterloo, Waterloo ON N2L 3G1, Canada.}% <-this % stops a space
\thanks{M Noguchi is with the Department of Kinesiology, University of Waterloo, Waterloo ON N2L 3G1, Canada.}
\thanks{TW Cleworth was with the Department of Kinesiology, University of Waterloo, Waterloo ON N2L 3G1, Canada, and is now with the School of Kinesiology and Health Science, York University, Toronto ON M3J 1P3, Canada.}}

\date{}

\begin{document}

\maketitle

\begin{abstract}
Postural instability is prevalent in aging and neurodegenerative disease, decreasing quality of life and independence.
Quantitatively monitoring balance control is important for assessing treatment efficacy and rehabilitation progress.
However, existing technologies for assessing postural sway are complex and expensive, limiting their widespread utility.
Here, we propose a monocular imaging system capable of assessing sub-millimeter 3D sway dynamics \added{during quiet standing}. \replaced{Two anatomical targets with known feature geometries were placed on the lumbar and shoulder. Upper and lower trunk 3D kinematic motion was automatically assessed from a set of 2D frames through geometric feature tracking and an inverse motion model.}{By physically embedding anatomical targets with known \textit{a priori} geometric models, 3D central and upper body kinematic motion was automatically assessed through geometric feature tracking and 3D kinematic motion inverse estimation from a set of 2D frames.}
Sway was tracked in 3D and compared between control and hypoperfusion conditions \added{in 14 healthy young adults}.
The proposed system demonstrated high agreement with a commercial motion capture system (error \replaced{$1.5 \times 10^{-4}~\text{mm}$, [$-0.52$, $0.52$]}{$4.4 \times 10^{-16} \pm 0.30$~mm, $r^2=0.9773$}).
\replaced{Between-condition differences in sway dynamics were observed in anterior-posterior sway during early and mid stance, and medial-lateral sway during mid stance}{Significant differences in sway dynamics were observed in early stance \replaced{lower trunk}{central} anterior-posterior sway (control: $147.1 \pm 7.43$~mm, hypoperfusion: $177.8 \pm 15.3$~mm; $p=0.039$) and mid stance \replaced{upper trunk}{upper body} coronal sway (control: $106.3 \pm 5.80$~mm, hypoperfusion: $128.1 \pm 18.4$~mm; $p=0.040$)} commensurate with \added{decreased} cerebral \deleted{blood flow (CBF)} perfusion\deleted{ deficit}, followed by recovered sway dynamics during late stance with \replaced{cerebral perfusion}{CBF} recovery.
This inexpensive single-camera system enables quantitative 3D sway monitoring for assessing neuromuscular balance control in weakly constrained environments.
\end{abstract}

\section{Introduction}
%\todo{cerebral blood flow for balance generally, application to falls; here studying cardiovascular and balance together}

Postural control is crucial for maintaining independence and quality of life. The two components of posture, orientation and balance, require continual adjustment and coordination between afferent sensory inputs and neuromuscular control~\cite{kandel2000neuralscience,winter1995balance,bell1998biomechanics}. Aging and neurodegenerative diseases (e.g., Parkinson's disease, multiple sclerosis) can cause deterioration in postural control, which is associated with increased risk of falls, and thus decreased quality of life~\cite{muir2010balancerisk,lin2012balance,rubenstein2006falls}. Dual-task, attention, and cortical recording paradigms have demonstrated the involvement of higher cortical centers during balance control~\cite{woollacott2002attention}\replaced{, supporting the notion that underlying mechanisms to age-related balance decline may include decreased cortical function.}{Therefore, potential underlying mechanisms to age-related balance decline may include decreased cortical function from low cerebral perfusion.}

%Falls occur in one of every three older adults, and account for 85\% of injury-related hospitalizations in adults over the age of 65~\cite{centers2014wisqars,tinetti1997falls}.
\added{Instability and falls in older adults is a complex multi-factorial problem, with risk factors such as muscle control, visual impairment and effects from prescription medication~\cite{tinetti1986falls,tromp2001fallrisk}.} One underlying mechanism for instability and falls in older adults is low cerebral blood flow (CBF)\added{, and thus cerebral hypoperfusion,} from impaired cardio- or cerebrovascular regulation~\cite{edlow2010posture,mehagnouls2000posture,gutkin2016orthostatic}. Two factors commonly affected by underlying disease pathophysiology are cerebral perfusion pressure and arterial partial pressure of carbon dioxide (P$_a$CO$_2$)~\cite{meng2015regulation}. Cerebral autoregulation maintains relatively constant CBF across a range of perfusion pressures, but rapid changes in arterial blood pressure, such as during a supine to stand transition, can result in an acute reduction of cerebral perfusion pressure, and subsequent reduction in blood flow to the brain.  This reduction in perfusion causes a decrease in energy metabolism and activity in the brain and central nervous system~\cite{sokoloff1981cortical}, which may affect cardio-postural balance control~\cite{goswami2017orthostatic}. There is a clinical need to objectively monitor posture and balance control for assessing treatment efficacy and rehabilitation~\cite{fortin2011clinical}.

Balance control \added{during quiet standing} has been largely investigated by measuring variation in center of pressure (CoP) using baropodometric platforms and center of mass (CoM) using 3D camera analysis. Laboratory grade \added{camera-based} measurement technologies have traditionally been restricted to assessment in controlled environments and to validate novel technologies, but are often too expensive or cumbersome to incorporate within clinical settings. Camera-based systems have traditionally been used to assess CoM sway. Specifically, marker-based motion capture systems have been widely used for estimating and tracking CoM during \added{quiet} standing\deleted{ and locomotion}, in which participants are fitted with retroreflective or actively illuminated markers and 3D kinematic data of body segments are tracked by a multi-camera laboratory setup. Although these systems are able to assess whole body motion, system expense, setup burden, and technical expertise have limited their clinical utility~\cite{nardone2010clinicalassess,visser2008posturography}. Less expensive multimodal alternatives using an iPad and 3D camera with retroreflective markers have been proposed for static posture assessment during lying posture~\cite{agustsson2019ipad3d}, but currently lack the ability to track dynamic sway in standing. Markerless technologies, such as multi-camera voxel reconstruction~\cite{wang2010voxel} and Microsoft Kinect-based technologies~\cite{webster2014kinect,yeung2014kinect}, have the benefit of ambient imaging without markers, but \replaced{existing methods have not demonstrated sub-millimeter accuracy at anatomically relevant locations during quiet standing and require specialized imaging setups}{have demonstrated insufficient accuracy for clinical utility}. Nevertheless, computer vision solutions for human pose estimation have shown tremendous promise in other motion-based applications, such as activity and gesture recognition. Although applications have largely been restricted to estimation in 2D space, integrating \textit{a priori} kinematic models with camera parameters has shown strong performance in 3D anatomical tracking of the hand joints~\cite{mueller2018ganerated} and arm~\cite{goncalves1995monocular}. The ability to accurately capture and track postural change dynamics associated with falls-related factors (such as CBF in aging) using low cost technical advances would be clinically useful for identifying an individual's potential falls risk beyond subjective assessment.

%Several currently available measurement devices are used to capture balance related behavior.  In clinical settings, wearable technologies with relatively high spatial resolution such as inertial measurement units (IMUs) or angular velocity transducers have been used to capture behaviour during more freely moving tasks~\cite{horak2015wearables}. In addition, low cost devices with poorer resolution such as the Wii Balance Board (Nintendo) and Kinect (Microsoft) have also been used to quantify movement during balance tasks~\cite{weaver2017wii,webster2014kinect}. Although these measurement units have proven useful across a wide range of tasks, there remains a need to develop techniques to quantify postural stability with high accuracy (sub-millimeter) using more readily available equipment for use in clinical, long-term care and naturalistic settings. The ability to accurately capture and track postural changes associated with falls-related factors (such as CBF in aging) using low cost technical advances would be clinically useful for identifying potential falls risk beyond subjective assessment.

In this paper, we propose a monocular 3D motion tracking imaging system for assessing sub-millimeter 3D sway dynamics \added{during quiet standing}. This system was designed to enable postural assessment in clinical or naturalistic environments. \replaced{Traditional motion tracking systems require unnecessarily complex whole-room configuration for sway tracking. This burden was alleviated through a one-time camera calibration and affixing physical targets with known unique \textit{a priori} geometries (``geometric target models'') on anatomically relevant locations. Using targets with known geometric models enabled high spatial accuracy estimation across the set of 2D frames by fitting a kinematic target motion model to the data. Thus, the camera can be affixed in any position and orientation as long as the targets are within the field of view. By calibrating to the scene orientation, upper and lower trunk sway coordinates were tracked in relevant biomechanical axes (anterior-posterior, medial-lateral, superior-inferior)  during quiet standing.}
{Using a one-time camera calibration procedure, 3D temporal \replaced{upper and lower trunk}{upper body and central} sway coordinates were assessed using a single-view (monocular) camera by fitting a kinematic model to automatically track salient target features with known \textit{a priori} geometric target models. These \textit{a priori} models enable higher spatial accuracy estimation compared to unsupervised 3D estimation methods by fitting a kinematic target motion model to the data. Data were transformed from the camera coordinate system into an anatomical Euclidean space to isolate sway in relevant biomechanical axes (anterior-posterior, medial-lateral, superior-inferior) by inferring absolute scene orientation. Using the lumbar kinematic matrix, a virtual \replaced{lower trunk}{central} sway coordinate was projected into the body to track \replaced{lower trunk}{central} sway alongside upper \replaced{trunk}{body} sway at the shoulder.}
This two-factor model (\replaced{lower and upper trunk}{central and upper body}) was used \replaced{to assess sway characteristics in anatomical planes using a repeated measures design with young healthy adults. For each participant, sway was monitored at baseline and under compromised cerebral perfusion (hypoperfusion), induced by pre-stand guided hyperventilation, to investigate sway dynamics related to cerebral perfusion levels.}{to assess sway characteristics in normal and compromised cerebral perfusion (hypoperfusion) cases by tracking sway characteristics in anatomical planes to investigate.}

\section{Methods}
\subsection{Data Collection \added{and Experimental Design}}
Fourteen young healthy adults (9/5 male/female, age 24.7~$\pm$~4.3, mass 74.3~$\pm$~11.7~kg) free from a history of cardiovascular, neurological and musculoskeletal disorders completed testing. Participants were instructed to refrain from caffeine and food consumption 2~hours prior to testing, and alcohol and strenuous exercise 24~hours before the laboratory visit. Study protocols and procedures were approved by a University of Waterloo Research Ethics Committee and conformed to the Declaration of Helsinki (ORE~19831). All participants provided written informed consent before testing procedures.

\added{Data was collected in the RIA Research Apartment, which is a purpose-built apartment to simulate realistic retirement and long-term care living conditions}. At the start of each testing session, anthropomorphic data were collected. Participants were then instrumented with an electrocardiogram to measure heart rate (Pilot 9200; Colin Medical Instruments, San Antonio, TX, USA), continuous arterial blood pressure finger plethysmograph to estimate cardiac stroke volume via Modelflow (Finapres Pro; FMS, Amsterdam, The Netherlands), transcranial Doppler ultrasound (WAKIe; Atys Medical, Soucieu en Jarrest, France), which insonated the right middle cerebral artery to estimate CBF, and a nasal cannula connected to a capnograph to measure \petco~(CD-3A CO$_2$ Analyzer, AMETEK Inc., Pittsburgh, PA, USA). Additionally, a spatially resolved near infrared spectroscopy probe (Portalite; Artinis Medical Systems, Elst, The Netherlands) was placed on the forehead above the right eye brow to measure cerebral tissue oxygenation index (TSI). Arterial blood pressure, stroke volume (SV), ECG, \petco, and cerebral blood flow velocity (CBFv) were recorded at 1000 Hz (PowerLab, LabChart, version 7.3.7; ADInstruments, Colorado Springs, CO). After instrumentation, participants assumed a supine position for 10~min before finger blood pressure was calibrated.
%using the ipsilateral brachial return-to-flow method. 

\replaced{Participants completed two stands in pseudorandomized order}{Participants were pseudorandomized to complete two repeated trials}: (1) supine to stand transition followed by 60~s of quiet standing, and (2) 2~min voluntary guided hyperventilation (20~breaths/min), immediately followed by a supine to stand transition and 60~s of quiet standing \added{with resumed normal breathing}. \added{The stands were repeated once, and data were averaged for each type of stand.} In accordance with the French Posturology Association guidelines, participants were instructed to stand without shoes, with their heels 2~cm apart, and feet angled at 30$^\circ$~\cite{gagey1988normes85}. Participants were instructed to keep their eyes closed and arms crossed to eliminate afferent visual feedback and reduce anticipatory arm movement~\cite{patla2002anticipatory}. During hyperventilation, participants were coached on depth of breathing to attain a drop of at least 10~mmHg end-tidal PCO$_2$ (\petco), which was used as a proxy for P$_a$CO$_2$. \added{Participants completed a practice stand prior to data collection, and all four stands were completed in the same session to avoid between-day sway variations~\cite{lovecchio2017swayrepeatability}.} %\todo{add that they stood onto a Wii Board if it doesn't make things tricky with Reviewr 1 comments}
%After each transition, participants were asked to rate subjective symptoms of (include symptom list here), and each transition was separated by 5~min of quiet, supine rest.

\subsection{\replaced{Monocular}{Single-View} 3D Sway Tracking}
\begin{figure*}
    \centering
    \includegraphics[width=0.9\textwidth]{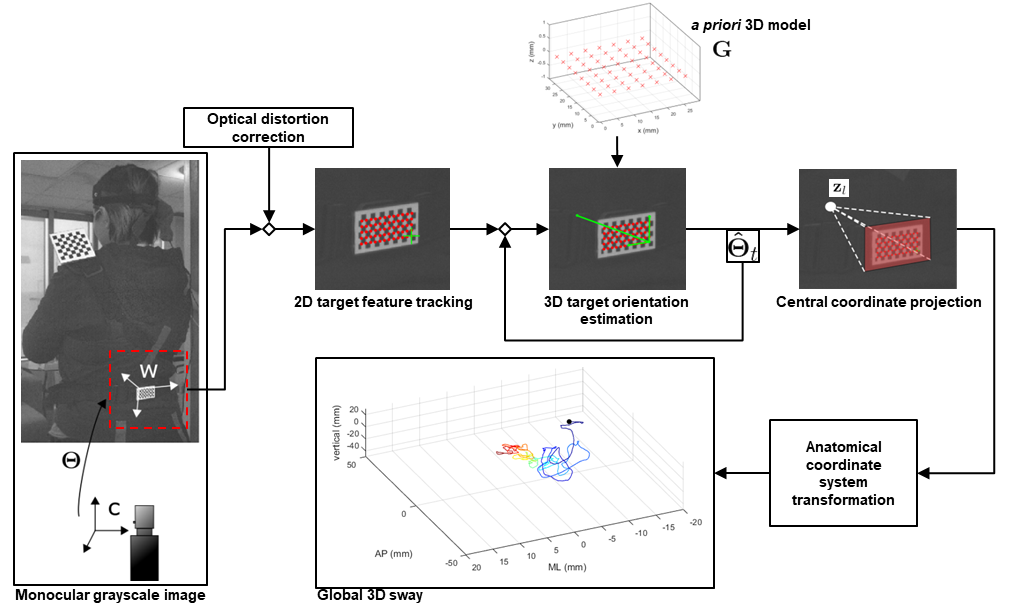}
    \caption{Overview of the monocular 3D sway estimation imaging system. Frames were captured posterior to the participant. Target features were tracked in 2D, and using \textit{a priori} 3D geometric model, 3D sway coordinates were estimated and transformed into anatomical space. \replaced{Upper and lower trunk}{Upper body and lumbar} sway coordinates were \added{estimated by tracking shoulder and lumbar targets} \deleted{tracked}, resulting in a global 3D sway profile (blue: early stance, red: late stance).}
    \label{fig:system_diagram}
\end{figure*}

The main goal was to develop a monocular imaging system for tracking 3D sway characteristics to assess balance control in weakly constrained (non-laboratory) environments. The problem was posed as a single-view \added{(monocular)} \textit{a priori} geometric model and inverse kinematic estimation problem with embedded anatomical target models. Fig.~\ref{fig:system_diagram} depicts an overview of the imaging system. A temporal sequence of 3D sway coordinates $\vec{z}_i \in \mathbb{R}^3$ was sought from a single sequence of 2D frames, where sway is represented across the anterior-posterior (AP), medial-lateral (ML), and superior-inferior (SI) axes. Given a video from a single posterior-facing camera, model feature coordinates from a shoulder and lumbar target were automatically tracked in 2D calibrated camera space (Section~\ref{sec:tracking}). Then, using known \textit{a priori} target model geometries, we fit a kinematic model to estimate the absolute sway position in 3D camera space (Section~\ref{sec:modelfit}). Finally, we projected the lumbar kinematic orientation into the center of the body to track a virtual \replaced{lower trunk}{central} sway coordinate. These coordinates were transformed into anatomical Euclidean space described by the AP, ML, SI axes (Section~\ref{sec:anatomicalspace}).

\subsubsection{Model Feature Tracking}
\label{sec:tracking}
We adopted a two-segment hinged biomechanical model of motion, with hinging effects between \replaced{lower and upper trunk}{central and upper body}. To separate sway from these two components, unique anatomical targets were affixed to the left shoulder and lumbar. \added{The targets were secured using adjustable torso harnesses with rigid attachment points positioned at the acromion process and L3 vertebra, identified through bony landmarks.} Since differences in balance control result in \replaced{lower trunk}{central} sway differences on the order of millimeters~\cite{gage2004invertedpendulum}, 3D tracking estimation was guided by \textit{a priori} geometric models to increase 3D estimation accuracy. The mathematical formulation presented here is generalizable to asymmetric target models with known root-relative feature coordinates. This asymmetry guarantees an orientation-dependent unique mapping onto 2D image space.

A single monocular grayscale camera (GS3-U3-41C6NIR, FLIR) was positioned 1~m behind the participant. Due to the high accuracy requirements of the system, optical distortions were estimated once and removed frame-by-frame using a two-step global-local camera calibration procedure~\cite{zhang2000calibration}. Images of a planar checkerboard pattern were recorded, and intrinsic and extrinsic camera parameters were modeled as a linear projection from 3D world coordinates to 2D image coordinates:
\begin{equation}
    \alpha~ {^o}\mat{X} = \mat{K}~{^c_w\mat{M}} ~{^w}\mat{X}
\end{equation}
where $\alpha$ is an arbitrary scale parameter, $^o\mat{X}$ and $^w\mat{X}$ are the checkerboard corner coordinates in the image plane and world coordinate system respectively, ${^c_w\mat{M}}$ is the extrinsic transformation matrix from 3D world to 3D camera coordinates, and $K$ is the intrinsic camera matrix:
\begin{equation}
    \mat{K} = \begin{bmatrix}
        f_x & s & x_0 \\
        0 & f_y & y_0 \\
        0 & 0 & 1
    \end{bmatrix}
\label{eq:K}
\end{equation}
where $(f_x,f_y)$ is the focal length, $s$ is skew, and $(x_0,y_0)$ is the principle point in the image plane. This matrix is fixed for the camera, and will be used later for estimating sway target positions. From this, we can define the world-to-image projection transformation function:
\begin{equation}
\label{eq:worldToImage}
    \Pi({^w}\mat{X}) = \frac{1}{\alpha} \mat{K}~{^c_w\mat{M}} ~{^w}\mat{X}
\end{equation}

The optical field distortion was estimated by refining the closed-form solution using nonlinear least squares minimization of a two-coefficient radial distortion~\cite{zhang2000calibration,heikkila1997}. This parameterization was used to undistort each frame prior to spatial processing to guarantee distance-independent homogeneous pixel spacing, and thus accurate sway tracking across the field of view.

In this study, we designed an asymmetric target model with equally spaced locally salient features. This is described by the \textit{a priori} feature model geometry matrix $\mat{G} \in \mathbb{R}^{n \times 3}$, which consists of 3D coordinates in world space \added{(i.e., the space defined by the target coordinate system)} and will be used for kinematic model fitting in Section~\ref{sec:modelfit}. Feature point coordinates $\vec{p}_{i} \in \mathbb{R}^2$ were automatically detected using multi-orientation kernel convolution with non-maxima suppression and sub-pixel localization~\cite{geiger2012corner}. Specifically, an interest point likelihood map was computed by convolving four feature kernels with the frame, and per-pixel feature likelihood was calculated by the maximum response over all prototype combinations. Sub-pixel feature localization was accomplished by solving a gradient orthogonality minimization problem:
\begin{equation}
    \vec{p}_i = \argmin_{\vec{q}_i} \sum_{\vec{n}_j \in \mathcal{N}(\vec{q}_i)}(\nabla_{\vec{q}_i}^T(\vec{n}_j-\vec{q}_i'))^2
\end{equation}
where $\vec{q}_i$ is a feature coordinate candidate, $\mathcal{N}(\vec{q}_i)$ and $\nabla_{\vec{q}_i}$ are the pixel neighborhood and image gradient at point $\vec{q}_i$ respectively, and $\vec{n}_j$ is a neighboring pixel. Thus, $\mat{P}=\{\vec{p}_i\}$ describes the set of feature coordinates after undergoing optical projection onto the image plane according to the camera intrinsics $\mat{K}$. The (unknown) 3D orientation of the geometric model was estimated by fitting a kinematic model to these data, which is discussed next.

\subsubsection{Kinematic Model Fitting}
\label{sec:modelfit}
Given the set of 2D feature coordinate predictions $\mat{P}$, we fit a kinematic motion model of the \textit{a priori} geometric model $\mat{G}$ to these data\added{, where $\mat{G}$ is a set of 3D feature coordinates in world space (see Fig.~\ref{fig:system_diagram})}. The kinematic model was designed to model the non-deformable nature of sway in free space with a fixed base of support. The model was parameterized by $\Theta = (\vec{t},\vec{R})$, where $\vec{t} \in \mathbb{R}^3$ is 3D translation, and $\vec{R} \in \mathbb{R}^3$ are the Euler angles describing 3D orientation. The optimal kinematic transformation, parameterized by these six degrees of freedom, was found by transforming the \textit{a priori} geometric model into the image plane using the calibrated camera model, and seeking a least squares fit to the feature prediction data:
\begin{equation}
    \hat{\Theta} = \argmin_\Theta \sum_i || \Pi(\mat{G}_i(\Theta)) - \vec{p}_i ||_2^2
\end{equation}
where $\Theta$ is the set of kinematic motion parameters, $\Pi$ is the projection transformation from 3D world coordinates to the 2D image plane from Eq.~(\ref{eq:worldToImage}), $\mat{G}_i(\Theta)$ are the transformed 3D coordinates of point $i$ from the geometric model $G$, and $\vec{p}_i$ is the feature prediction in image space.

This problem was solved using a two-step approach, consisting of an initializing and refinement step. To motivate this approach, we note that sway dynamics during quiet standing exhibit small and relatively smooth changes between each time point. In the first frame, we initialized the parameters $\Theta$ using a closed form planar estimation solution of the camera extrinsics~\cite{zhang2000}. In subsequent frames, noting that frame-to-frame sway differences are generally small (sub-millimeter), we set the initial conditions for the current frame (at time $t_c$) to the previous frame kinematic parameters ($\hat{\Theta}_{t_c-dt}$), and computed the optimal fit using Levenberg-Marquardt non-linear least-squares minimization. This approach avoided potential erroneous fits in local minima in other parts of the energy field, and we empirically found it produced higher accuracy than randomly initialized iterative optimization.

This optimization was performed on both the shoulder and lumbar targets separately, using their respective geometry priors. The target origin $\vec{z}_u$ was used to track upper \replaced{trunk}{body} motion. The torso kinematic parameters were used to project a virtual coordinate 10~cm deep into the body, which was used to track \replaced{lower trunk}{central} motion:
\begin{equation}
    \vec{z}_l = M_\Theta \vec{\Delta}_l
\end{equation}
where $\vec{\Delta}_l$ is the torso vector in homogeneous world space coordinates, and $M_\Theta$ is the motion matrix parameterized by $\Theta \in \mathbb{R}^6$, described by Eq.~(\ref{eq:mtheta}).
\begin{figure*}[t!]
\normalsize
% Store the current equation number.
% Set the equation number to one less than the one
% desired for the first equation here.
% The value here will have to changed if equations
% are added or removed prior to the place these
% equations are referenced in the main text.
\begin{equation}
    M_{\Theta} = \begin{bmatrix}
        \cos\Theta_1\cos\Theta_2 & \cos\Theta_1\sin\Theta_2\sin\Theta_3-\sin\Theta_1\cos\Theta_3 & \cos\Theta_1\sin\Theta_2\cos\Theta_3+\sin\Theta_1\sin\Theta_3 & \Theta_4 \\
        \sin\Theta_1\cos\Theta_2 & \sin\Theta_1\sin\Theta_2\sin\Theta_3+\cos\Theta_1\cos\Theta_3 & \sin\Theta_1\sin\Theta_2\cos\Theta_3-\cos\Theta_1\sin\Theta_3 & \Theta_5 \\
        -\sin\Theta_2 & \cos\Theta_2\sin\Theta_3 & \cos\Theta_2\cos\Theta_3 & \Theta_6 \\
        0 & 0 & 0 & 1 
    \end{bmatrix}
    \label{eq:mtheta}
\end{equation}
% Restore the current equation number.
% IEEE uses as a separator
\hrulefill
% The spacer can be tweaked to stop underfull vboxes.
\vspace*{4pt}
\end{figure*}

\subsubsection{Anatomical Space Transformation}
\label{sec:anatomicalspace}
To analyze posture sway patterns in anatomically relevant space, the kinematic parameters $\Theta$ were transformed from camera coordinate system into an anatomical coordinate system described by 1D axes (AP, ML, and SI) and derivative 2D planes (sagittal, transverse and coronal). A forward-facing calibration board was positioned in the scene, and its extrinsic orientation matrix was estimated using the calibration procedure from Section~\ref{sec:tracking}. Denoting this matrix as $\mathcal{E}$, sway in anatomical space coordinates was computed as:
\begin{align}
    \vec{z}_u' &= \mathcal{E}^{-1}~ \vec{z}_u \\
    \vec{z}_l' &= \mathcal{E}^{-1}~ \vec{z}_l
\end{align}
where $\vec{z}_u'$ and $\vec{z}_l'$ are the \replaced{upper and lower trunk}{shoulder and central} sway coordinates in the anatomical coordinate system defined by $\mathcal{E}^{-1}$, the inverse of the planar target orientation in camera coordinates. The signals were denoised using a second order Savitzky-Golay filter~\cite{savitzkygolay} with 0.5~s time window, which empirically modeled the smooth nature of sway well.

\subsection{Data Analysis}
The interval between ECG R-waves was used to calculate heart rate (HR). Cardiac output (CO) was calculated as the product of HR and SV. Systolic blood pressure (SBP), diastolic blood pressure (DBP), and mean arterial pressure (MAP) were the respective maximum, minimum and mean arterial pressures within each cardiac cycle. Identical analysis was performed to determine systolic, diastolic, and mean cerebral blood flow velocity (CBFv).  \petco~was determined by identifying the peak CO$_2$ concentration at the end of each exhalation, and the concentration was then converted to partial pressure. TSI was recorded at 50 Hz (Oxysoft, version 3.0.95, Artinis, Medical Systems, Elst, The Netherlands) and was averaged into 1~s bins. Beat-by-beat cardiovascular and breath-by-breath \petco~data were linearly interpolated to 1~s time points, and subsequently time aligned with the cerebral oxygenation data for analysis. For all variables, supine baseline values were calculated as 30 s averages (from 45~s to 15~s before the posture transition). \added{Cardiovascular variables during e}arly, mid, and late stance \deleted{values} were calculated as averages during the first 10~s of stance time.

To compare sway variations in control and hypoperfusion conditions, \replaced{sway dynamics were assessed at three time bins during quiet standing according to expected cardiovascular response relative to upright stance time (t=0~s): initial cerebrovascular \replaced{decrease}{deficit} (``early stance'', 0--20~s), overshoot and recovery onset (``mid stance'', 20--40~s), and sustained recovery (``late stance'', 40--60~s).}{sway was divided and analyzed across three time bins spanning early, mid and late stance (0--20~s, 20--40~s, 40--60~s).} \added{The start of stand (0~s) was determined at the time when the participant's full weight was transferred onto a pressure platform following the initial downward force overshoot.} For each time bin, the total path length (TPL) in each anatomical axis (AP, ML, SI) and anatomical plane (transverse, sagittal, coronal) was computed as a summary metric for balance control~\cite{salavati2009tpl}:
\begin{equation}
    \mathcal{L}_\mathcal{A}(T) = \sum_{\tau_i \in T} \sqrt{(x_{i+1}-x_i)^2+(y_{i+1}-y_i)^2}
\end{equation}
where $(x_i,y_i)$ are projected coordinates in the anatomical plane $\mathcal{A}$, and $T$ is the stance time bin. $\mathcal{L}$ within a 1D anatomical axis (i.e., AP, ML, SI) was computed by setting $y_i=0$. This formulation is analogous to the average velocity magnitude during the time frame~\cite{clark2010wiiboard}. 

Two-way repeated measures ANOVA, with within-subject factors of condition (control vs. hypoperfused) and time (baseline, early, mid, late stance), was performed on physiological measures. Normality was confirmed by the Shapiro-Wilks test, as well as visual inspection by using histograms and q-q plots of the residual distributions for each variable. Post hoc analysis was performed using paired sample t-tests to test differences across conditions within each time bin, and non-parametric ANOVA for non-normal sway TPL data~\cite{brunner2002nparld, noguchi2012nparld}. The $p$-values were adjusted via Bonferroni correction for assessing statistical significance. \added{Between-participant variance was removed using Cousineau-Morey normalization for reporting descriptive statistics.} We reported \added{Cohen's $d$ effect size and} statistically significant results when $p<0.05$\deleted{, and trending results when $p<0.10$}. Data are presented as mean $\pm$ SEM.

% To account for inter-participant balance control variations, repeated measures analysis of variance (ANOVA) was used to statistically compare sway dynamics in the control and hypoperfusion conditions. Nonparametric analyses were performed for the sway data since the assumptions of sphericity and normality were not satisfied. A non-parametric ANOVA with a within-subject factor of ventilation condition (two levels) was performed using the nparLD 2.1 library (R3.3.0; R Development Core Team)~\cite{brunner2002nparld, noguchi2012nparld}. This analysis was chosen based on its robustness for datasets with small sample sizes and outliers. This analysis was repeated for each stance time bin (three levels) and Bonferroni correction was applied to investigate at which time points the differences between ventilation conditions occurred.

System accuracy was evaluated against a commercial active motion capture system with an accuracy of 0.1~mm and resolution of 0.01~mm (Optotrak, Northern Digital Inc, Canada). \added{The global coordinate system was calibrated to the standing position such that target positions were within the manufacturer's characterized measurement volume.} A subset of four participants (3/1 male/female) was used to \replaced{assess}{validate} system accuracy. Each participant stood quietly for 60~s across four trials to simulate different sway patterns: eyes open on foam, eyes closed on foam, eyes open on ground, eyes closed on ground\added{, totalling 16~unique stands}. During ``eyes open'' stance, the participants looked at a visual target approximately 3~m in front at eye level.  Three infrared emitting diodes were affixed to the same rigid bodies as the video camera targets. All motion tracking data were recorded simultaneously. Optotrak based kinematic data was sampled at 120 Hz, resampled to 30 Hz by linear interpolation to match the frame rate of the monocular kinematic system, and the coordinate system origins were aligned. \deleted{For the purpose of this study, only the data from shoulder markers in the AP direction are discussed (similar results were observed for lumbar markers and ML displacements).} Bland-Altman analysis was used to compare the two measurement systems. Specifically, all data were concatenated across participants for each of the kinematic systems, and the point-by-point differences were quantified through correlation and equality.

\section{Results}
Section~\ref{sec:results_accuracy} presents \replaced{the accuracy}{accuracy validation} of the monocular imaging system \replaced{compared to}{against} a gold standard motion capture system. Section~\ref{sec:results_sway} presents repeated measures analysis of sway characteristics in control versus cerebral hypoperfusion across the relevant time bins. Video~1 \added{(Supplementary Materials)} shows the integration of cardiovascular response and postural sway estimation during a postural transition.

\subsection{3D Estimation Accuracy}
\label{sec:results_accuracy}
\begin{figure*}
\centering
\includegraphics[width=0.8\textwidth]{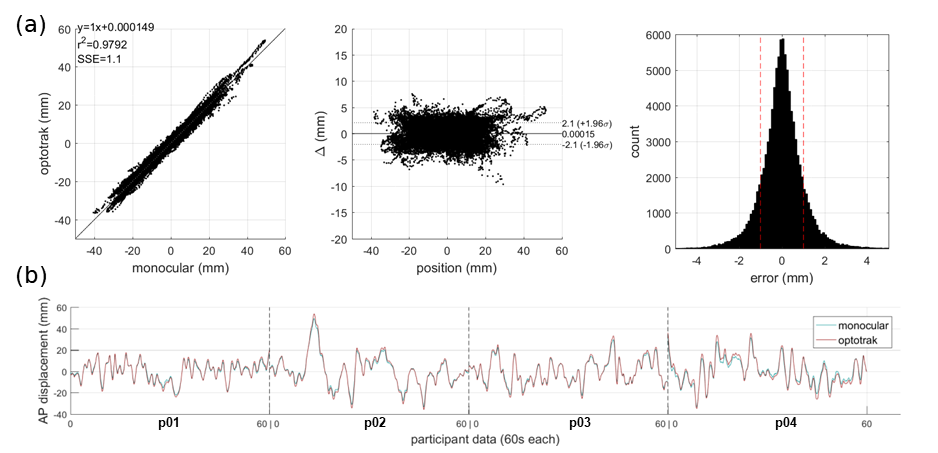}
\caption{Accuracy of the proposed monocular system compared to a whole-room motion capture system. (a) Bland-Altman analysis of systematic error shows strong agreement and sub-millimeter accuracy (error \replaced{$1.5 \times 10^{-4}~\text{mm}$, [$-0.52$, $0.52$]}{$4.4\times 10^{-16} \pm 0.30$~mm}). (b) Example monocular and Optotrak time series signals showing upper \replaced{trunk}{body} anterior-posterior sway during 60~s quiet stand with eyes closed on foam.}
\label{fig:results_optotrak}
\end{figure*}

Fig.~\ref{fig:results_optotrak} shows the agreement results between the proposed \added{monocular system} and \added{the gold standard} motion capture system. Bland-Altman analysis demonstrated no systematic error between the systems (error \replaced{$1.5 \times 10^{-4}~\text{mm}$, [$-0.52$, $0.52$]}{$4.4\times 10^{-16} \pm 0.30$~mm}), and no proportional error (\replaced{$y=1.00x + 1.49 \times 10^{-4}$, $r^2=0.9792$}{$y=1.01x+8.4\times 10^{-16}$, $r^2=0.9773$}). The equality line fell within the confidence interval of the mean difference. Thus, the monocular imaging system demonstrated comparable postural sway tracking results to a whole-room gold standard method during quiet standing tasks. 

\subsection{Postural Sway and Cardiovascular Response}
\label{sec:results_sway}

% \todo{FIGURE: change to normalized CI/SEM?}
\begin{figure*}
\centering
\includegraphics[width=0.9\textwidth]{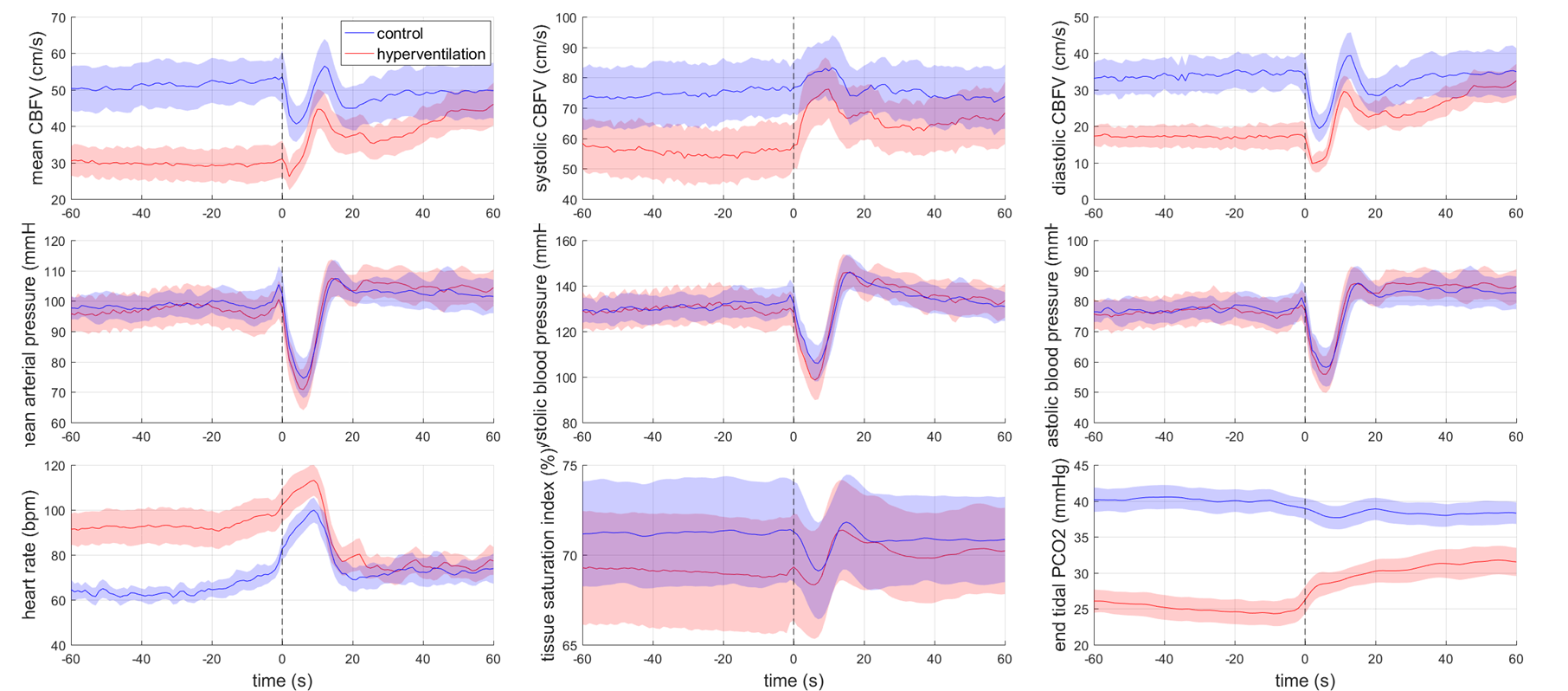}
\caption{Cardiovascular response to standing during normal and reduced cerebral perfusion (mean, standard error). Data were time-normalized based on established upright posture at t=0~s. Binned summary statistics are reported in Table~\ref{tab:cardiovasc}. (CBFv: cerebral blood flow velocity; PCO$_2$: partial pressure of carbon dioxide)}
\label{fig:cv_plots}
\end{figure*}

\begin{figure}
    \centering
    \includegraphics[width=0.5\textwidth]{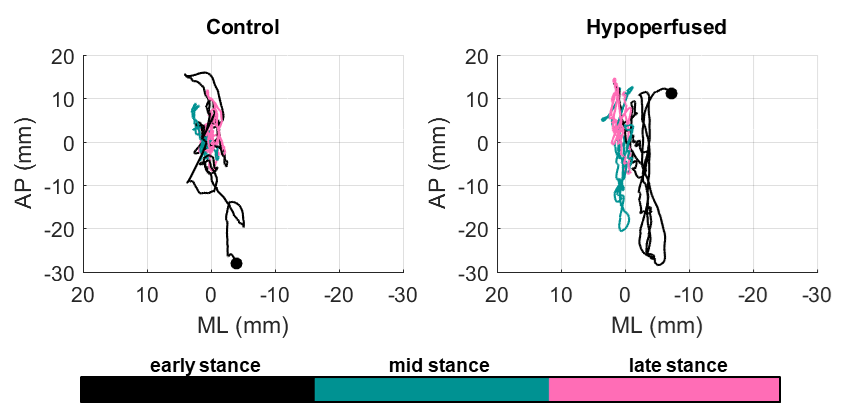}
    \caption{Example sway data of a participant with \added{decreased} cerebral blood flow \deleted{deficit} during hypoperfusion. AP-ML only is shown for visual clarity. Standing in a hypoperfused state caused larger early (\replaced{black}{blue}) and mid (\replaced{teal}{green}) stance sway dynamics compared to control. By late stance (\replaced{pink}{red}), sway stabilized in both conditions.}
    \label{fig:p14sway}
\end{figure}

\begin{figure*}
\centering
\includegraphics[width=0.9\textwidth]{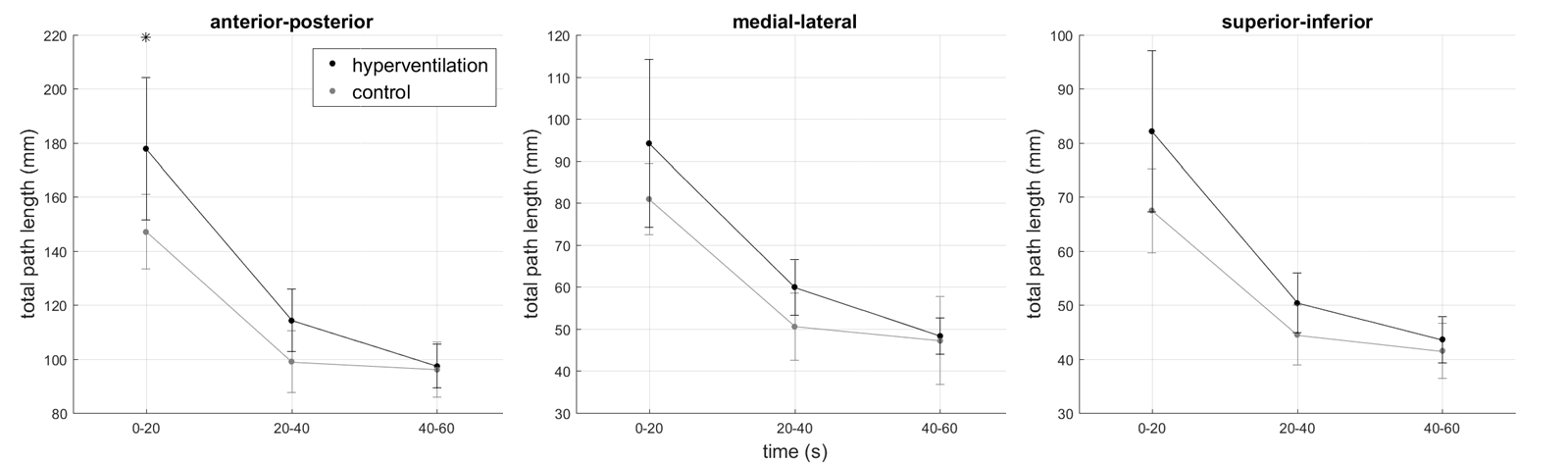}
\caption{Lower trunk 3D sway \added{(mean, 95\% confidence intervals)} binned across early, mid, and late stance times in \replaced{selected}{the} axes of anatomical motion. Similar results were found for upper trunk sway (see Table~\ref{tab:sway}). ($^*p<0.05$\deleted{, $^+p<0.10$})}
\label{fig:significance}
\end{figure*}

Both cardiovascular and sway showed the largest difference between control and hypoperfusion conditions during early stance, with gradual recovery to baseline by late stance, demonstrated by a significant main effect of time on all measures. \added{Normal respiratory rate ($11.5~\text{min}^{-1},~\text{SD}=2.9$) was successfully attained upon standing.} Fig.~\ref{fig:cv_plots} shows the primary time-synchronized cardiorespiratory and cerebrovascular responses to standing (at $t=0$) in both the control (blue) and hyperventilation (red) conditions. There were no significant main effects on perfusion condition in blood pressure measures (mean, diastolic, systolic), indicating preserved central arterial pressure across conditions. Significant Condition~$\times$~Time interaction terms were observed in all physiological measures, and are expanded and discussed below. Table~\ref{tab:cardiovasc} provides summary time-binned cardiovascular measures alongside statistical significance.

\subsubsection{Hyperventilation Caused Hypoperfusion}
Cerebral hypoperfusion was attained for each participant through hyperventilation-induced respiratory alkalosis. \petco~was significantly lower in the hyperventilation compared to the control conditions during all time points ($p<0.001$; see Table~\ref{tab:cardiovasc}). In the hypoperfusion condition, \petco~was a significant different across all stance times ($p<0.001$) except from mid to late stance ($p=0.43$). In the control condition, there were no significant differences in \petco~across time points.

In both perfusion conditions, all participants demonstrated vasopressor response to upright posture with a transient reduction in blood pressure from baseline to early stance ($p<0.001$), and compensatory increase in HR. There were no differences between conditions in systolic, diastolic, or mean blood pressure, hence between-condition differences in CBF and oxygenation were attributed to differences in \petco~from hyperventilation. CBFv variables (systolic, diastolic, mean) and TSI all had significant Condition~$\times$~Time interaction effects due the compensatory mechanisms of CBFv after the termination of hyperventilation and acclimation to upright posture.

Fig.~\ref{fig:p14sway} shows sway traces of a representative participant with \added{decreased} CBF \deleted{deficit} during hyperventilation, and demonstrates the primary effects of standing on postural control. During early stance (\replaced{black}{blue}), sway TPL increased in hypoperfused (248.3~mm) versus control (166.2~mm). At this time, arterial blood pressure, CBFv and TSI are transiently low due to active standing. During mid stance (\replaced{teal}{green}), the TPL difference between hypoperfused (151.9~mm) and control (113.2~mm) starts to diminish as CBF and perfusion start to recover. By late stance (\replaced{pink}{red}), balance control had been re-established (132.5 vs. 130.0~mm) owing to cerebral reperfusion and cardiovascular homeostasis. Whole-sample results binned by stance time are presented and discussed below. \added{Table~\ref{tab:sway} presents whole sample TPL results, and } Fig.~\ref{fig:significance} shows TPL distributions calculated across selected anatomical axes for each time bin.

\subsubsection{Early Stance}
Significant between-condition reductions in all CBFv and oxygenation variables were observed in both baseline supine ($p<0.001$) and early stance ($p<0.03$), indicating acute onset of hypoperfused state during hyperventilation. Within the hypoperfusion condition, no significant differences were observed from baseline to early stance in mean CBFv or TSI, indicating sustained impaired cerebrovascular perfusion and oxygenation during the initial stance phase. A concomitant statistically significant \added{($p=0.039$)} increase in \replaced{lower trunk}{central} \replaced{AP}{anterior-posterior} sway from control \deleted{($147.1 \pm 5.9$~mm)} to hypoperfusion \deleted{($177.8 \pm 11.1$~mm)} conditions was observed \replaced{(147.1~mm vs. 177.8~mm, $d=0.92$)}{($p=0.039$)}, as well as \deleted{trending} \added{large} differences in \replaced{upper trunk AP sway (191.2~mm vs. 232.0~mm, $d=0.93$), and AP-derivative planes of motion in both lower trunk (transverse: 183.3~mm vs. 221.1~mm, $d=0.89$; saggital: 165.0~mm vs. 200.3~mm, $d=0.91$) and upper trunk (transverse: 227.9~mm vs. 272.7~mm, $d=0.84$; saggital: 202.5~mm vs. 252.7~mm, $d=0.86$).}{transverse (control: $183.3 \pm 7.1$~mm; hypoperfusion: $221.0 \pm 14.3$~mm; $p=0.098$) and sagittal (control: $165.0 \pm 6.8$~mm; hypoperfusion: $200.3 \pm 13.1$~mm; $p=0.075$) planes.}

\subsubsection{Mid Stance}
During mid stance, significant reductions in all CBFv variables were observed ($p<0.02$), but TSI was no longer statistically different ($p=0.11$). Within the hypoperfusion condition, mean CBFv and TSI increased from early to mid stance ($p<0.001$), indicating initial cerebral perfusion recovery onset. \replaced{AP}{Anterior-posterior} sway decreased, showing \replaced{decreased magnitude and effect size compared to early stance in both lower trunk ($d=0.84$ vs. $0.92$) and upper trunk ($d=0.81$ vs. $0.93$)}{no significant trends in either \replaced{lower or upper trunk}{central or upper body} sway}. However, \added{there were large ML sway differences in both lower trunk (50.5~mm vs. 59.9~mm, $d=0.81$) and upper trunk (53.9~mm vs. 63.3~mm, $d=0.80$), and} \replaced{a statistically significant ($p=0.040$) increase was}{significant differences were} observed in upper \replaced{trunk}{body} coronal sway \added{(66.5~mm vs. 76.4~mm, $d=0.68$).}\deleted{(control: $66.5 \pm 4.1$~mm; hypoperfusion: $76.4 \pm 3.6$~mm; $p=0.040$), and \deleted{trending} differences in both upper \replaced{trunk}{body} (control: $53.9 \pm 3.5$~mm; hypoperfusion: $63.3 \pm 2.8$~mm; $p=0.052$) and \replaced{lower trunk}{central} (control: $50.5 \pm 3.4$~mm; hypoperfusion: $59.9 \pm 8.5$~mm; $p=0.055$) medial-lateral sway.}

\subsubsection{Late Stance}
During late stance, cerebral perfusion levels were largely recovered through no between-condition significant differences in diastolic CBFv or TSI, but mean CBFv and systolic CBFv remained low ($p<0.04$). Within the hypoperfusion condition, cerebral perfusion continued to recover, demonstrated by significant increases in mean CBFv and TSI compared to mid stance ($p<0.01$). There was no significant between-condition difference in any sway measures in late stance, \added{and effect sizes across all directions were small ($d<0.3$)}, indicating regained balance control following initial hypoperfusion onset.

\begin{table*}
\centering
\footnotesize
\caption{\replaced{\textsc{Cardiovascular Response Data for Each Condition Binned by Stance Time (mean $\pm$ SEM).}}{Summary statistics of cardiovascular data binned by stance time (mean $\pm$ SEM). Arterial blood pressure variables (MAP, SBP, DBP) did not show significant main effect differences on perfusion condition. All other variables showed significant main and interaction effects, indicating impaired cerebral blood flow and perfusion during hyperventilation. (HR: heart rate; MAP/SBP/DBP: mean/systolic/diastolic arterial blood pressure; SV: stroke volume; CO: cardiac output; CBFv: cerebral blood flow velocity; TSI: tissue saturation index; \petco: end-tidal PCO$_2$)}}
\label{tab:cardiovasc}
\begin{tabular}{ccccc}
\hline
\multicolumn{1}{c|}{}                                & \multicolumn{2}{c|}{\textbf{Supine Baseline}}                                                                  & \multicolumn{2}{c}{\textbf{Early Stance}}                                                                     \\
\multicolumn{1}{c|}{}                                & Control                               & \multicolumn{1}{c|}{Hypoperfusion}                         & Control                                & \multicolumn{1}{c}{Hypoperfusion}                         \\ \hline
\multicolumn{1}{l|}{HR (bpm)}                        & 63.1 $\pm$ 1.6 $^b$                      & \multicolumn{1}{c|}{92.2 $\pm$ 4.0 $^{* b, c, d}$}                & 92.8 $\pm$ 2.6 $^{a, c, d}$                 & \multicolumn{1}{c}{108.5 $\pm$ 3.8 \(^{* a, c, d}\)}               \\
\multicolumn{1}{l|}{MAP (mmHg)}                      & 98.6 $\pm$ 2.1 $^b$                      & \multicolumn{1}{c|}{97.8 $\pm$ 3.1 $^{b, c}$}                     & 83.1 $\pm$ 3.1  $^{a, c, d}$                & \multicolumn{1}{c}{79.2 $\pm$ 3.2 $^{a, c, d}$}                  \\
\multicolumn{1}{l|}{SBP (mmHg)}                      & 130.6 $\pm$ 3.0 $^{b}$                    & \multicolumn{1}{c|}{130.8 $\pm$ 4.1 $^{b, c}$}                    & 114.6 $\pm$ 3.6 $^{a, c, d}$               & \multicolumn{1}{c}{108.3 $\pm$ 4.0 $^{a, c, d}$}                 \\
\multicolumn{1}{l|}{DBP (mmHg)}                      & 77.3 $\pm$ 1.9 $^{b, c, d}$                & \multicolumn{1}{c|}{77.1 $\pm$ 2.6 $^{b, c, d}$}                   & 64.4 $\pm$ 3.1 $^{a, c, d}$                & \multicolumn{1}{c}{62.2 $\pm$ 3.0 $^{a, c, d}$}                  \\
\multicolumn{1}{l|}{SV (mL)}                         & 92.8 $\pm$ 4.4 $^{d}$                     & \multicolumn{1}{c|}{90.9 $\pm$ 3.8 $^{b, c, d}$}                  & 88.9 $\pm$ 4.1 $^{d}$                      & \multicolumn{1}{c}{79.9 $\pm$ 4.3 $^{* a}$}                      \\
\multicolumn{1}{l|}{CO (L/min)}                      & 5.8 $\pm$ 0.3 $^{b}$                       & \multicolumn{1}{c|}{8.4 $\pm$ 0.6 $^{* c, d}$}                     & 8.2 $\pm$ 0.5 $^{a, c, d}$                  & \multicolumn{1}{c}{8.6 $\pm$ 0.6 $^{c, d}$}                       \\
\multicolumn{1}{l|}{Mean CBFv (cm/s)}                & 51.4 $\pm$ 3.2                       & \multicolumn{1}{c|}{29.5 $\pm$ 2.3 $^{* c, d}$}                    & 46.2 $\pm$ 2.7                         & \multicolumn{1}{c}{32.2 $\pm$ 2.2 $^{* c, d}$}                    \\
\multicolumn{1}{l|}{Sys CBFv (cm/s)}                 & 74.6 $\pm$ 5.2                       & \multicolumn{1}{c|}{54.9 $\pm$ 4.6 $^{* b, c, d}$}                & 80.2 $\pm$ 4.6                        & \multicolumn{1}{c}{69.0 $\pm$ 4.8 $^{* a}$}                      \\
\multicolumn{1}{l|}{Dia CBFv (cm/s)}                 & 34.1 $\pm$ 2.4 $^{b}$                      & \multicolumn{1}{c|}{17.1 $\pm$ 1.5 $^{* b, c, d}$}                 & 25.9 $\pm$ 2.0 $^{a, d}$                    & \multicolumn{1}{c}{14.2 $\pm$ 1.5 $^{* a, c, d}$}                 \\
\multicolumn{1}{l|}{TSI (\%)}                        & 73.2 $\pm$ 1.0 $^{b}$                      & \multicolumn{1}{c|}{70.7 $\pm$ 1.1 $^{*}$ }                        & 71.3 $\pm$ 1.3 $^{a}$                       & \multicolumn{1}{c}{69.6 $\pm$ 1.5 $^{* c}$}                       \\
\multicolumn{1}{l|}{\petco~(mmHg)}                   & 40.2 $\pm$ 0.8                        & \multicolumn{1}{c|}{24.9 $\pm$ 0.9 $^{* b, c, d}$}                 & 38.1 $\pm$ 0.8                         & \multicolumn{1}{c}{28.0 $\pm$ 0.8 $^{* a, c, d}$}                 \\ \hline \hline
\multicolumn{1}{c|}{}                                & \multicolumn{2}{c|}{\textbf{Mid Stance}}                                                                  & \multicolumn{2}{c}{\textbf{Late Stance}}                                                                     \\
\multicolumn{1}{c|}{}                                & Control                               & \multicolumn{1}{c|}{Hypoperfusion}                         & Control                                & \multicolumn{1}{c}{Hypoperfusion}                         \\ \hline
\multicolumn{1}{l|}{HR (bpm)}                        & 70.2 $\pm$ 2.6 $^{b}$                      & \multicolumn{1}{c|}{76.3 $\pm$ 2.8 $^{a, b}$}                     & 72.7 $\pm$ 2.9 $^{b}$                      & \multicolumn{1}{c}{75.1 $\pm$ 2.6 $^{a, b}$}                      \\
\multicolumn{1}{l|}{MAP (mmHg)}                      & 103.1 $\pm$ 2.0 $^{b}$                     & \multicolumn{1}{c|}{105.6 $\pm$ 2.5 $^{a, b}$}                     & 102.8 $\pm$ 2.9 $^{b}$                     & \multicolumn{1}{c}{104.2 $\pm$ 2.9 $^{b}$}                       \\
\multicolumn{1}{l|}{SBP (mmHg)}                      & 139.7 $\pm$ 2.5 $^{b}$                     & \multicolumn{1}{c|}{141.0 $\pm$ 3.3 $^{a, b, d}$}                 & 134.1 $\pm$ 3.3 $^{b}$                     & \multicolumn{1}{c}{134.6 $\pm$ 3.8 $^{b, c}$}                    \\
\multicolumn{1}{l|}{DBP (mmHg)}                      & 82.4 $\pm$ 2.1 $^{a, b}$                   & \multicolumn{1}{c|}{84.9 $\pm$ 2.1 $^{a, b}$}                      & 83.6 $\pm$ 2.8 $^{a, b}$                   & \multicolumn{1}{c}{85.1 $\pm$ 2.4 $^{a, b}$}                      \\
\multicolumn{1}{l|}{SV (mL)}                         & 88.4 $\pm$ 5.7 $^{d}$                     & \multicolumn{1}{c|}{83.7 $\pm$ 4.4 $^{a, d}$}                     & 78.1 $\pm$ 4.8 $^{a, b, c}$                & \multicolumn{1}{c}{75.2 $\pm$ 4.4 $^{a, c}$}                     \\
\multicolumn{1}{l|}{CO (L/min)}                      & 6.2 $\pm$ 0.4 $^{b, d}$                    & \multicolumn{1}{c|}{6.3 $\pm$ 0.4 $^{a, b, d}$}                    & 5.7 $\pm$ 0.4 $^{b, c}$                     & \multicolumn{1}{c}{5.6 $\pm$ 0.4 $^{a, b, c}$}                    \\
\multicolumn{1}{l|}{Mean CBFv (cm/s)}                & 46.5 $\pm$ 3.1                       & \multicolumn{1}{c|}{36.7 $\pm$ 2.4 $^{* a, b, d}$}                 & 49.3 $\pm$ 3.6                        & \multicolumn{1}{c}{42.1 $\pm$ 2.5 $^{* a, b, c}$}                 \\
\multicolumn{1}{l|}{Sys CBFv (cm/s)}                 & 76.4 $\pm$ 4.4                      & \multicolumn{1}{c|}{65.7 $\pm$ 4.5 $^{* a}$}                      & 73.7 $\pm$ 5.0                        & \multicolumn{1}{c}{65.2 $\pm$ 4.6 $^{* a}$}                      \\
\multicolumn{1}{l|}{Dia CBFv (cm/s)}                 & 30.2 $\pm$ 2.5 $^{d}$                      & \multicolumn{1}{c|}{23.2 $\pm$ 1.9 $^{* a, b, d}$}                 & 34.2 $\pm$ 3.1 $^{b, c}$                   & \multicolumn{1}{c}{28.7 $\pm$ 2.0 $^{a, b, c}$}                   \\
\multicolumn{1}{l|}{TSI (\%)}                        & 72.0 $\pm$ 1.2 $^{d}$                      & \multicolumn{1}{c|}{71.2 $\pm$ 1.4 $^{b, d}$}                      & 71.7 $\pm$ 1.2 $^{c}$                       & \multicolumn{1}{c}{70.5 $\pm$ 1.4 $^{c}$}                         \\
\multicolumn{1}{l|}{\petco~(mmHg)}                   & 38.5 $\pm$ 0.8                        & \multicolumn{1}{c|}{30.3 $\pm$ 0.8 $^{* a, b}$}                    & 38.1 $\pm$ 0.8                         & \multicolumn{1}{c}{31.3 $\pm$ 1.1 $^{* a, b}$}                    \\ \hline
\multicolumn{5}{l}{\begin{tabular}[c]{@{}l@{}}* = significantly different from control condition value at a given time point (post-hoc analysis)\\
$^{a, b, c, d}$ = within-condition significantly different from baseline, early, mid, or late stance value, respectively. \\
HR: heart rate; MAP/SBP/DBP: mean/systolic/diastolic arterial blood pressure; SV: stroke volume; \\ CO: cardiac output; CBFv: cerebral blood flow velocity; TSI: tissue saturation index; \petco: end-tidal PCO$_2$)\end{tabular}}
\end{tabular}
\end{table*}

\begin{table*}
\centering
\footnotesize
\caption{\textsc{Sway Total Path Length (mm) Binned by Direction and Stance Time (mean $\pm$ SEM).}}
\label{tab:sway}

\begin{tabular}{ll|ccc|ccc}
              &             & \multicolumn{3}{c|}{\textbf{Lower Trunk}}                                  & \multicolumn{3}{c}{\textbf{Upper Trunk}}                                  \\
              &             & \multicolumn{1}{c}{\textbf{Control}} & \multicolumn{1}{c}{\textbf{Hyperventilation}} & \multicolumn{1}{c|}{\textbf{$d$}} & \multicolumn{1}{c}{\textbf{Control}} & \multicolumn{1}{c}{\textbf{Hyperventilation}} & \multicolumn{1}{c}{\textbf{$d$}} \\ \hline
\textbf{AP}   & \textbf{T1} & 147.1 $\pm$ 5.9 & 177.8 $\pm$ 11.1                              & 0.92                                   & 191.2 $\pm$ 8.3                      & 232.9 $\pm$ 14.7                              & 0.93                                   \\
\textbf{}     & \textbf{T2} & 99.0 $\pm$ 4.8                                              & 114.3 $\pm$ 4.9                               & 0.84                                   & 129.0 $\pm$ 7.5                      & 147.7 $\pm$ 4.5                               & 0.81                                   \\
\textbf{}     & \textbf{T3} & 96.1 $\pm$ 4.3                                              & 97.4 $\pm$ 3.4                                & 0.09                                   & 126.6 $\pm$ 6.4                      & 131.3 $\pm$ 5.1                               & 0.22                                   \\ \hline
\textbf{ML}   & \textbf{T1} & 80.9 $\pm$ 3.6                                              & 94.2 $\pm$ 8.5                                & 0.55                                   & 90.3 $\pm$ 3.1                       & 98.6 $\pm$ 9.0                                & 0.33                                   \\
\textbf{}     & \textbf{T2} & 50.5 $\pm$ 3.4                                              & 59.9 $\pm$ 2.8                                & 0.81                                   & 53.9 $\pm$ 3.5                       & 63.3 $\pm$ 2.8                                & 0.8                                    \\
\textbf{}     & \textbf{T3} & 47.2 $\pm$ 4.4                                              & 48.3 $\pm$ 1.8                                & 0.09                                   & 51.1 $\pm$ 5.0                       & 52.3 $\pm$ 1.9                                & 0.08                                   \\ \hline
\textbf{SI}   & \textbf{T1} & 67.4 $\pm$ 3.3                                              & 82.1 $\pm$ 6.3                                & 0.78                                   & 41.1 $\pm$ 3.0                       & 57.7 $\pm$ 10.9                               & 0.55                                   \\
\textbf{}     & \textbf{T2} & 44.5 $\pm$ 2.3                                              & 50.4 $\pm$ 2.3                                & 0.67                                   & 29.9 $\pm$ 2.7                       & 31.2 $\pm$ 2.3                                & 0.14                                   \\
\textbf{}     & \textbf{T3} & 41.5 $\pm$ 2.2                                              & 43.6 $\pm$ 1.8                                & 0.28                                   & 29.1 $\pm$ 1.8                       & 26.7 $\pm$ 3.6                                & -0.22                                  \\ \hline
\textbf{APML} & \textbf{T1} & 183.3 $\pm$ 7.1 & 221.1 $\pm$ 14.3                              & 0.89                                   & 227.9 $\pm$ 8.8                      & 272.7 $\pm$ 18.1                              & 0.84                                   \\
\textbf{}     & \textbf{T2} & 120.8 $\pm$ 5.8                                             & 141.0 $\pm$ 5.6                               & 0.94                                   & 149.6 $\pm$ 8.2                      & 172.8 $\pm$ 5.11                              & 0.91                                   \\
\textbf{}     & \textbf{T3} & 116.2 $\pm$ 6.7                                             & 118.5 $\pm$ 3.7                               & 0.12                                   & 145.5 $\pm$ 8.8                      & 150.9 $\pm$ 4.9                               & 0.2                                    \\ \hline
\textbf{APSI} & \textbf{T1} & 165.0 $\pm$ 6.8 & 200.3 $\pm$ 13.1                              & 0.91                                   & 202.5 $\pm$ 9.2                      & 252.7 $\pm$ 20.0                              & 0.86                                   \\
\textbf{}     & \textbf{T2} & 110.8 $\pm$ 5.4                                             & 127.3 $\pm$ 5.5                               & 0.82                                   & 137.2 $\pm$ 8.1                      & 155.9 $\pm$ 5.0                               & 0.74                                   \\
\textbf{}     & \textbf{T3} & 106.9 $\pm$ 4.8                                             & 108.8 $\pm$ 3.9                               & 0.12                                   & 134.6 $\pm$ 6.5                      & 138.0 $\pm$ 6.3                               & 0.14                                   \\ \hline
\textbf{MLSI} & \textbf{T1} & 116.6 $\pm$ 4.9 & 139.4 $\pm$ 10.7                              & 0.73                                   & 106.3 $\pm$ 4.3                      & 128.2 $\pm$ 13.4                              & 0.59                                   \\
\textbf{}     & \textbf{T2} & 74.7 $\pm$ 4.1                                              & 86.8 $\pm$ 3.7                                & 0.83                                   & 66.5 $\pm$ 4.1                       & 76.4 $\pm$ 3.6                                & 0.68                                   \\
\textbf{}     & \textbf{T3} & 69.7 $\pm$ 5.2                                              & 72.3 $\pm$ 2.5                                & 0.17                                   & 63.8 $\pm$ 5.1                       & 63.4 $\pm$ 3.9                                & -0.02                                  \\ \hline
\multicolumn{8}{l}{AP: anterior-posterior; ML: medial-lateral; SI: superior-inferior;} \\
\multicolumn{8}{l}{T1: early stance; T2: mid stance; T3: late stance $d$: Cohen's $d$}
\end{tabular}
\end{table*}

\section{Discussion}
\added{In this study we proposed a novel 3D monocular imaging system for monitoring postural sway in weakly constrained imaging environments. The system distinguished between upper and lower trunk movement using two unique wearable targets with \textit{a priori} geometric models, yielding a two-segment kinematic model assuming hinging effects between upper and lower trunk segments. 
Lower spinal displacement has been shown to accurately estimate CoM changes during gait tasks~\cite{yang2014sacral}. Additional anatomical targets may provide enhanced dynamic CoM measurement at the expense of increased physical setup.}

\added{Balance control involves neuromuscular coordination to maintain the CoM within the base of support~\cite{ku2014balanceamputeereview,maki1996postural} and stabilization following intrinsic or extrinsic disturbances~\cite{horak2006swayfalls}.
It has been commonly assumed that the role of CoP is to correct CoM deviations through neuromuscular control of ankle and hip torque~\cite{winter1998stiffness}. Recent studies have suggested that the link between CoP and CoM may be more complex~\cite{carpenter2010exploratorycop}, and play complementary roles in balance control. The prevalence of baropodometric CoP monitoring may be due to the ease of setup and technology acquisition. However, 3D body motion may be able to provide complementary indicators of neuromuscular insufficiency in addition to baropodometric monitoring. Motion capture systems are often used for assessing segmental and CoM displacement, but are expensive and require complex setups compared to baropodometric platforms. Thus, a less expensive portable imaging system may allow for new CoP-CoM co-analysis for balance assessment in naturalistic environments.}

\added{A hyperventilation protocol was used to modulate cerebral perfusion through hypocapnia-induced cerebral vasoconstriction resulting in relative cerelbral hypoxia~\cite{sakellari1997hypervent}. In healthy older adults, impaired cerebral vasoreactivity, but not impaired cerebral autoregulation, is associated with increased falls risk~\cite{sorond2010cerebrovascular} primarily in the form of orthostatic intolerance (OI)~\cite{gupta2007oh}. Similarly, responses to head up tilt in OI groups following parabolic flight have been linked to cerebral vasoconstriction and not to systemic hypotension~\cite{serrador2000parabolic}. Similar responses have been observed in classic OI population, specifically increased heart rate, decreased CBFv, and increased cerebrovascular resistance~\cite{novak1998hypocapnia}. These manipulations demonstrate similar effects to traditional OI, and thus appear to be effective proxies for studying imbalance in older adults.}

\added{Differences in sway were commensurate with physiological changes in cerebral perfusion. Baroreflex response modulates heart rate through orthostatic reduction of blood pressure during posture transition~\cite{rowell1993}. Baroreflex sensitivity decreases with age~\cite{gribbin1971baroreflex,kornet2005baroreflex}, and has been linked to autonomic dysfunction, including orthostatic hypotension~\cite{low2015ohmechanisms}. Antihypertensive treatment for older adults living needs to assess risk for orthostatic hypotension and consider the risk for cerebral hypoperfusion~\cite{finucane2017oh}. Significant between-condition differences in early and mid stance were observed in CBFv and not blood pressure measures, with concomitant differences in sway magnitude. Cerebrovascular autoregulation~\cite{sorond2010cerebrovascular} promoted recovery of CBFv in late stance, which was reflected by recovered sway dynamics. Thus, combining blood pressure data, or direct measures of CBFv or cerebral oxygenation, with functional measures of balance control may increase diagnostic aid and treatment efficacy in individuals with orthostatic hypotension.}

\added{Differences were observed for both lower and upper trunk sway kinematics during the hypoperfused state support a multi-joint model of motion. 
%There has been a wide range of body segmentation for assessing quiet standing, ranging from total body center of mass to individual rigid body segment (e.g., 14~segment bilateral model~\cite{gage2004invertedpendulum}). The current study evaluated a two-segment kinematic model, assuming hinging effects between upper and \replaced{lower}{central} trunk segments. Expanded models could be used to estimate dynamic CoM. 
Traditional body segment analysis requires placement of many optical markers on the body, and reconstruction of body segments using a multi-camera setup. The proposed system alleviates the setup load by distinguishing between upper and lower trunk motion through two individual unique targets, which may reduce the barrier to adoption in clinical settings. Furthermore, since the imaging system tracks posterior anatomical markers, no facial information is recorded or required for 3D sway analysis, and participant or patient privacy can be maintained. Thus, if paired with baropodometric assessment, sway motion can be additionally assessed while maintaining patient privacy. This may be beneficial in home care and health care environments where privacy is an important factor in technology adoption~\cite{arning2015camera}.}

\added{The study was designed as a repeated measures study with pseudorandomized stand ordering to minimize confounding factors such as order and demographic effects. The manipulation of increased respiration to reduce CBFv was stopped prior to standing so should not have affected postural movement, and self-reported symptoms of dizzyness and light-headedness suggested that the predominant factor in increased sway was cerebral hypoperfusion; this is commensurate with previous findings~\cite{sakellari1997hypervent,sakellari1997hyperventilation2}. Our model was able to achieve aging-like hypoperfusion in healthy young adults. Further investigations are needed to evaluate the monocular imaging system with individuals with chronically impaired balance control. Furthermore, additional investigations are required to evaluate a formal linear relationship between target motion and multi-segmental CoM analysis~\cite{gage2004invertedpendulum}.}

% The study was designed as a repeated measures study with pseudorandomized stand ordering to minimize confounding factors such as order, demographic and pharmaceutical effects. Hyperventilation was used to attain cerebral hypoperfusion, followed by resumed normal breathing during the stand. Respiratory fatigue may contribute to increased sway, however self-reported symptoms of dizzyness and light-headedness suggest that the predominant factor in increased sway is cerebral hypoperfusion, and is commensurate with previous findings~\cite{sakellari1997hypervent,sakellari1997hyperventilation2}. \todo{show these in Results} Repeated measures design enabled direct comparison of hyperventilation condition on postural sway while keeping other factors constant.

\deleted{In this study we proposed a novel 3D monocular imaging system for monitoring postural sway in weakly constrained imaging environments. The system distinguished between \replaced{upper and lower trunk}{upper body and central} sway using two unique wearable targets with \textit{a priori} geometric models. \replaced{Sway accuracy was assessed using}{The sway was validated against} a gold standard multi-camera whole room motion capture system during different postural tasks. Sway patterns linked to orthostatic hypotension were assessed using a hyperventilation protocol to acutely reduce cerebral perfusion prior to standing. Results showed significant differences in \replaced{upper and lower trunk}{upper body and central} sway patterns during early and mid stance between perfusion conditions, followed by similar sway patterns during late stance as cerebral perfusion recovered.}

\deleted{We observed significant differences in sway commensurate with physiological changes in CBF and perfusion. Hyperventilation caused hypocapnia (reduced blood carbon dioxide) from forced exhalation, which results in vasoconstriction in cerebral vessels from the reduced partial pressure of CO$_2$, and ultimately cerebral hypoxia~\cite{sakellari1997hypervent}. Compensatory systemic effects were observed to compensate for the hypoxic condition. In healthy older adults, impaired cerebral vasoreactivity, but not impaired cerebral autoregulation, is associated with increased falls risk~\cite{sorond2010cerebrovascular} primarily in the form of orthostatic intolerance (OI)~\cite{gupta2007oh}. Similarly, responses to head up tilt in OI groups following parabolic flight have been linked to cerebral vasoconstriction and not to systemic hypotension~\cite{serrador2000parabolic}. Similar responses have been observed in classic OI population, namely increased heart rate, decreased CBFv, and increased cerebrovascular resistance~\cite{novak1998hypocapnia}. These manipulations demonstrate similar effects to traditional OI, and thus appear to be effective proxies for studying imbalance in older adults.}

\deleted{The increased heart rate observed while transferring from a supine to standing position was likely an effect of the acute gravitationally-driven drop in arterial blood pressure triggering the baroreflex~\cite{rowell1993}. The baroreflex response increases heart rate and total peripheral resistance in an attempt to maintain adequate arterial blood pressure. This effect was observed, where arterial pressures were well maintained across conditions. Baroreflex sensitivity decreases with age~\cite{gribbin1971baroreflex,kornet2005baroreflex}, and has been linked to autonomic dysfunction, including orthostatic hypotension~\cite{low2015ohmechanisms}. Current antihypertensive treatment for older adults living with orthostatic hypotension is governed by observations of decreased postural blood pressure~\cite{james1999baroreflex}. Combining blood pressure data with functional measures of balance control may increase diagnostic aid and treatment efficacy.}

\deleted{Decreases in CBFv rather than blood pressure may provide better indications for postural instability and cerebral perfusion recovery. Whereas SBP dropped during the initial stages of standing, systolic CBFv was well maintained. Decreased systolic CBFv has been previously observed during pre-syncope~\cite{thomas2009ioh}. The drops in mean and diastolic arterial blood pressure on standing were accompanied by reductions in CBFv that could impair O$_2$ delivery. Intact cerebrovascular autoregulation~\cite{sorond2010cerebrovascular} promoted rapid recovery of CBFv associated with re-establishment of postural control by late stance following cerebral hypoperfusion.}
%Maintained systolic CBFv were observed in this study, which may be the cause for re-established postural control by late stand following cerebral hypoperfusion.

% indicate a deviation from the rigid inverted pendulum model~\cite{winter1998stiffness,gage2004invertedpendulum}, and rather

\deleted{Differences were observed for both \replaced{lower and upper trunk}{central and upper body} sway kinematics during the hypoperfused state support a multi-joint model of motion. Specifically, between condition differences during early stance were observed in \replaced{lower trunk}{central} sway (anterior-posterior, transverse, sagittal), and mid stance were observed in both \replaced{lower}{central} (medial-lateral) and \replaced{upper trunk}{upper body} (medial-lateral, coronal). Further studies are needed to expand these differences related to impaired neuromuscular control. There has been a wide range of body segmentation for assessing quiet standing, ranging from total body center of mass to individual rigid body segment (e.g., 14~segment bilateral model~\cite{gage2004invertedpendulum}). The current study evaluated a two-segment kinematic model, assuming hinging effects between upper and \replaced{lower}{central} body segments. \added{Expanded models could be used to estimate dynamic CoM.} Traditional body segment analysis requires placement of many optical markers on the body, and reconstruction of body segments using a multi-camera setup. The proposed system alleviates the setup load by distinguishing between \replaced{upper and lower trunk}{upper and central body} motion through two individual targets, which may reduce the barrier to adoption in clinical settings. Furthermore, since the imaging system tracks posterior anatomical markers, no facial information is recorded or required for 3D sway analysis, and thus participant or patient privacy can be maintained. Thus, if paired with baropodometric assessment, sway motion can be additionally assessed while maintaining patient privacy. This may be beneficial in home care and health care environments where privacy is an important factor in technology adoption~\cite{arning2015camera}.}

% differences in early stand anterior-posterior, transverse, and saggital, as well as mid-stand medial-lateral central sway were observed, whereas mid-stand medial-lateral and coronal upper body sway were observed.

% It has been shown that the vestibular system modulates vasodilation and cerebral blood flow~\cite{serrador2009vestibular} \todo{so what?}

%talk about time 2 ML sway

%tsi no increase from early-mid, but sig increase mid-late and early-late. cbfv increase early-mid. first increase source (cbfv) leads to perfusion (tsi) later?

% sys and mean CBFv were still low during late stand. would it regain upon muscle pump activation?

% "changes in steady‐state cerebral hemodynamics between postures do not appear to have a large functional consequence on the dynamic regulation of CBF." https://www.ncbi.nlm.nih.gov/pmc/articles/PMC5328778/

\section{Conclusion}
In this paper, we proposed a novel monocular kinematic imaging system for assessing 3D postural sway during \added{quiet standing following} postural transition under varying cerebral perfusion levels. By physically embedding geometric priors, \replaced{lower and upper trunk}{central and upper body} kinematic motion was automatically tracked through feature tracking and 3D orientation inverse estimation. \replaced{Lower trunk}{Central body} sway was estimated by forward projecting a virtual coordinate \added{from the lumbar} midway through the body, and transforming the data into an anatomical coordinate system. \replaced{System accuracy was assessed using}{The system was validated against} a commercial motion capture system and demonstrated sub-millimeter accuracy across different types of sway. Hypocapnia-induced cerebral hypoperfusion showed increases in sway total path length in \replaced{anterior-posterior motion}{\replaced{lower trunk}{central} anterior-posterior, transverse and sagittal motion} during early \added{and mid} stance, as well as increases in medial-lateral and coronal sway during mid stance. No differences were found during late stance, suggesting recovered cerebral perfusion and neuromuscular control. This system provides \replaced{cost effective}{inexpensive}, accurate quantitative postural sway tracking in weakly constrained (non-laboratory) environments\added{, and may be used} as a screening tool for cerebrovascular sufficiency and balance control in resource constrained settings.

\section{Acknowledgments}
The authors are grateful to Dr. Laura Fitzgibbon-Collins for her assistance with protocol development.

\bibliographystyle{IEEEtran}
\bibliography{bibliography}

% Generated by IEEEtran.bst, version: 1.14 (2015/08/26)
\begin{thebibliography}{10}
\providecommand{\url}[1]{#1}
\csname url@samestyle\endcsname
\providecommand{\newblock}{\relax}
\providecommand{\bibinfo}[2]{#2}
\providecommand{\BIBentrySTDinterwordspacing}{\spaceskip=0pt\relax}
\providecommand{\BIBentryALTinterwordstretchfactor}{4}
\providecommand{\BIBentryALTinterwordspacing}{\spaceskip=\fontdimen2\font plus
\BIBentryALTinterwordstretchfactor\fontdimen3\font minus
  \fontdimen4\font\relax}
\providecommand{\BIBforeignlanguage}[2]{{%
\expandafter\ifx\csname l@#1\endcsname\relax
\typeout{** WARNING: IEEEtran.bst: No hyphenation pattern has been}%
\typeout{** loaded for the language `#1'. Using the pattern for}%
\typeout{** the default language instead.}%
\else
\language=\csname l@#1\endcsname
\fi
#2}}
\providecommand{\BIBdecl}{\relax}
\BIBdecl

\bibitem{kandel2000neuralscience}
E.~R. Kandel, J.~H. Schwartz, T.~M. Jessell, S.~Siegelbaum, and A.~Hudspeth,
  \emph{{Principles of Neural Science}}, 5th~ed.\hskip 1em plus 0.5em minus
  0.4em\relax New York: McGraw-Hill, 2012.

\bibitem{winter1995balance}
D.~A. Winter, ``Human balance and posture control during standing and
  walking,'' \emph{Gait \& Posture}, vol.~3, no.~4, pp. 193--214, 1995.

\bibitem{bell1998biomechanics}
F.~Bell, \emph{Principles of Mechanics and Biomechanics}.\hskip 1em plus 0.5em
  minus 0.4em\relax Nelson Thornes, 1998.

\bibitem{muir2010balancerisk}
S.~W. Muir, K.~Berg, B.~Chesworth, N.~Klar, and M.~Speechley, ``Quantifying the
  magnitude of risk for balance impairment on falls in community-dwelling older
  adults: a systematic review and meta-analysis,'' \emph{Journal of Clinical
  Epidemiology}, vol.~63, no.~4, pp. 389--406, 2010.

\bibitem{lin2012balance}
H.~W. Lin and N.~Bhattacharyya, ``Balance disorders in the elderly:
  epidemiology and functional impact,'' \emph{The Laryngoscope}, vol. 122,
  no.~8, pp. 1858--1861, 2012.

\bibitem{rubenstein2006falls}
L.~Z. Rubenstein, ``Falls in older people: epidemiology, risk factors and
  strategies for prevention,'' \emph{Age and Ageing}, vol.~35, no. suppl\_2,
  pp. ii37--ii41, 2006.

\bibitem{woollacott2002attention}
M.~Woollacott and A.~Shumway-Cook, ``Attention and the control of posture and
  gait: a review of an emerging area of research,'' \emph{Gait \& Posture},
  vol.~16, no.~1, pp. 1--14, 2002.

\bibitem{tinetti1986falls}
M.~E. Tinetti, T.~F. Williams, and R.~Mayewski, ``Fall risk index for elderly
  patients based on number of chronic disabilities,'' \emph{The American
  Journal of Medicine}, vol.~80, no.~3, pp. 429--434, 1986.

\bibitem{tromp2001fallrisk}
A.~Tromp, S.~Pluijm, J.~Smit, D.~Deeg, L.~Bouter, and P.~Lips, ``Fall-risk
  screening test: a prospective study on predictors for falls in
  community-dwelling elderly,'' \emph{Journal of Clinical Epidemiology},
  vol.~54, no.~8, pp. 837--844, 2001.

\bibitem{edlow2010posture}
B.~L. Edlow, M.~N. Kim, T.~Durduran, C.~Zhou, M.~E. Putt, A.~G. Yodh, J.~H.
  Greenberg, and J.~A. Detre, ``The effects of healthy aging on cerebral
  hemodynamic responses to posture change,'' \emph{Physiological Measurement},
  vol.~31, no.~4, p. 477, 2010.

\bibitem{mehagnouls2000posture}
D.~J. Mehagnoul-Schipper, L.~C. Vloet, W.~N. Colier, W.~H. Hoefnagels, and
  R.~W. Jansen, ``Cerebral oxygenation declines in healthy elderly subjects in
  response to assuming the upright position,'' \emph{Stroke}, vol.~31, no.~7,
  pp. 1615--1620, 2000.

\bibitem{gutkin2016orthostatic}
M.~Gutkin and J.~M. Stewart, ``Orthostatic circulatory disorders: from nosology
  to nuts and bolts,'' \emph{American Journal of Hypertension}, vol.~29, no.~9,
  pp. 1009--1019, 2016.

\bibitem{meng2015regulation}
L.~Meng and A.~W. Gelb, ``Regulation of cerebral autoregulation by carbon
  dioxide,'' \emph{Anesthesiology: The Journal of the American Society of
  Anesthesiologists}, vol. 122, no.~1, pp. 196--205, 2015.

\bibitem{sokoloff1981cortical}
L.~Sokoloff, ``Relationships among local functional activity, energy
  metabolism, and blood flow in the central nervous system.'' in
  \emph{Federation Proceedings}, vol.~40, no.~8, 1981, pp. 2311--2316.

\bibitem{goswami2017orthostatic}
N.~Goswami, A.~P. Blaber, H.~Hinghofer-Szalkay, and J.-P. Montani,
  ``Orthostatic intolerance in older persons: etiology and countermeasures,''
  \emph{Frontiers in Physiology}, vol.~8, p. 803, 2017.

\bibitem{fortin2011clinical}
C.~Fortin, D.~Ehrmann~Feldman, F.~Cheriet, and H.~Labelle, ``Clinical methods
  for quantifying body segment posture: a literature review,'' \emph{Disability
  and Rehabilitation}, vol.~33, no.~5, pp. 367--383, 2011.

\bibitem{nardone2010clinicalassess}
A.~Nardone and M.~Schieppati, ``The role of instrumental assessment of balance
  in clinical decision making.'' \emph{European Journal of Physical and
  Rehabilitation Medicine}, vol.~46, no.~2, pp. 221--237.

\bibitem{visser2008posturography}
J.~E. Visser, M.~G. Carpenter, H.~van~der Kooij, and B.~R. Bloem, ``The
  clinical utility of posturography,'' \emph{Clinical Neurophysiology}, vol.
  119, no.~11, pp. 2424--2436, 2008.

\bibitem{agustsson2019ipad3d}
A.~Agustsson, M.~Gislason, P.~Ingvarsson, E.~Rodby-Bousquet, and T.~Sveinsson,
  ``Validity and reliability of an {iPad} with a three-dimensional camera for
  posture imaging,'' \emph{Gait \& Posture}, vol.~68, pp. 357--362, 2019.

\bibitem{wang2010voxel}
F.~Wang, M.~Skubic, C.~Abbott, and J.~M. Keller, ``Body sway measurement for
  fall risk assessment using inexpensive webcams,'' in \emph{2010 Annual
  International Conference of the IEEE Engineering in Medicine and Biology},
  2010, pp. 2225--2229.

\bibitem{webster2014kinect}
D.~Webster and O.~Celik, ``Systematic review of kinect applications in elderly
  care and stroke rehabilitation,'' \emph{Journal of NeuroEngineering and
  Rehabilitation}, vol.~11, no.~1, p. 108, 2014.

\bibitem{yeung2014kinect}
L.~Yeung, K.~C. Cheng, C.~Fong, W.~C. Lee, and K.-Y. Tong, ``Evaluation of the
  {Microsoft Kinect} as a clinical assessment tool of body sway,'' \emph{Gait
  \& Posture}, vol.~40, no.~4, pp. 532--538, 2014.

\bibitem{mueller2018ganerated}
F.~Mueller, F.~Bernard, O.~Sotnychenko, D.~Mehta, S.~Sridhar, D.~Casas, and
  C.~Theobalt, ``{GANerated} hands for real-time {3D} hand tracking from
  monocular {RGB},'' in \emph{Proceedings of the IEEE Conference on Computer
  Vision and Pattern Recognition}, 2018, pp. 49--59.

\bibitem{goncalves1995monocular}
L.~Goncalves, E.~Di~Bernardo, E.~Ursella, and P.~Perona, ``Monocular tracking
  of the human arm in {3D},'' in \emph{Proceedings of IEEE International
  Conference on Computer Vision}, 1995, pp. 764--770.

\bibitem{gagey1988normes85}
P.~Gagey, R.~Gentaz, J.~Guillamon, G.~Bizzo, C.~Bodot-Brégard, C.~Debruille,
  and C.~Baudry, \emph{Normes 85}, 2nd~ed.\hskip 1em plus 0.5em minus
  0.4em\relax Paris: Association Francaise de Posturologie, 1988.

\bibitem{patla2002anticipatory}
A.~E. Patla, M.~G. Ishac, and D.~A. Winter, ``Anticipatory control of center of
  mass and joint stability during voluntary arm movement from a standing
  posture: interplay between active and passive control,'' \emph{Experimental
  Brain Research}, vol. 143, no.~3, pp. 318--327, 2002.

\bibitem{lovecchio2017swayrepeatability}
N.~Lovecchio, M.~Zago, L.~Perucca, and C.~Sforza, ``Short-term repeatability of
  stabilometric assessments,'' \emph{Journal of motor behavior}, vol.~49,
  no.~2, pp. 123--128, 2017.

\bibitem{gage2004invertedpendulum}
W.~H. Gage, D.~A. Winter, J.~S. Frank, and A.~L. Adkin, ``Kinematic and kinetic
  validity of the inverted pendulum model in quiet standing,'' \emph{Gait \&
  Posture}, vol.~19, no.~2, pp. 124--132, 2004.

\bibitem{zhang2000calibration}
Z.~Zhang, ``A flexible new technique for camera calibration,'' \emph{IEEE
  Transactions on Pattern Analysis and Machine Intelligence}, vol.~22, 2000.

\bibitem{heikkila1997}
J.~Heikkila and O.~Silven, ``A four-step camera calibration procedure with
  implicit image correction,'' in \emph{Proc. Computer Vision and Pattern
  Recognition}, 1997, pp. 1106--1112.

\bibitem{geiger2012corner}
A.~Geiger, F.~Moosmann, O.~Car, and B.~Schuster, ``Automatic camera and range
  sensor calibration using a single shot,'' in \emph{2012 IEEE International
  Conference on Robotics and Automation}, 2012, pp. 3936--3943.

\bibitem{zhang2000}
Z.~Zhang, ``A flexible new technique for camera calibration,'' \emph{IEEE
  Transactions on Pattern Analysis and Machine Intelligence}, vol.~22, 2000.

\bibitem{savitzkygolay}
A.~Savitzky and M.~J. Golay, ``Smoothing and differentiation of data by
  simplified least squares procedures.'' \emph{Analytical Chemistry}, vol.~36,
  no.~8, pp. 1627--1639, 1964.

\bibitem{salavati2009tpl}
M.~Salavati, M.~R. Hadian, M.~Mazaheri, H.~Negahban, I.~Ebrahimi, S.~Talebian,
  A.~H. Jafari, M.~A. Sanjari, S.~M. Sohani, and M.~Parnianpour, ``Test-retest
  reliability of center of pressure measures of postural stability during quiet
  standing in a group with musculoskeletal disorders consisting of low back
  pain, anterior cruciate ligament injury and functional ankle instability,''
  \emph{Gait \& Posture}, vol.~29, no.~3, pp. 460--464, 2009.

\bibitem{clark2010wiiboard}
R.~A. Clark, A.~L. Bryant, Y.~Pua, P.~McCrory, K.~Bennell, and M.~Hunt,
  ``{Validity and reliability of the Nintendo Wii balance board for assessment
  of standing balance},'' \emph{Gait \& Posture}, vol.~31, no.~3, pp. 307--310,
  2010.

\bibitem{brunner2002nparld}
E.~Brunner, S.~Domhof, and F.~Langer, \emph{Nonparametric analysis of
  longitudinal data in factorial experiments}.\hskip 1em plus 0.5em minus
  0.4em\relax Wiley-Interscience, 2002, vol. 406.

\bibitem{noguchi2012nparld}
K.~Noguchi, Y.~R. Gel, E.~Brunner, and F.~Konietschke, ``{nparLD}: an {R}
  software package for the nonparametric analysis of longitudinal data in
  factorial experiments,'' \emph{Journal of Statistical Software}, vol.~50,
  no.~12, 2012.

\bibitem{yang2014sacral}
F.~Yang and Y.-C. Pai, ``Can sacral marker approximate center of mass during
  gait and slip-fall recovery among community-dwelling older adults?''
  \emph{Journal of Biomechanics}, vol.~47, no.~16, pp. 3807--3812, 2014.

\bibitem{ku2014balanceamputeereview}
P.~X. Ku, N.~A.~A. Osman, and W.~A. B.~W. Abas, ``Balance control in lower
  extremity amputees during quiet standing: a systematic review,'' \emph{Gait
  \& Posture}, vol.~39, no.~2, pp. 672--682, 2014.

\bibitem{maki1996postural}
B.~E. Maki and W.~E. McIlroy, ``Postural control in the older adult,''
  \emph{Clinics in Geriatric Medicine}, vol.~12, no.~4, pp. 635--658, 1996.

\bibitem{horak2006swayfalls}
F.~B. Horak, ``Postural orientation and equilibrium: what do we need to know
  about neural control of balance to prevent falls?'' \emph{Age and Ageing},
  vol.~35, no. suppl\_2, pp. ii7--ii11, 2006.

\bibitem{winter1998stiffness}
D.~A. Winter, A.~E. Patla, F.~Prince, M.~Ishac, and K.~Gielo-Perczak,
  ``Stiffness control of balance in quiet standing,'' \emph{Journal of
  Neurophysiology}, vol.~80, no.~3, pp. 1211--1221, 1998.

\bibitem{carpenter2010exploratorycop}
M.~Carpenter, C.~Murnaghan, and J.~Inglis, ``Shifting the balance: evidence of
  an exploratory role for postural sway,'' \emph{Neuroscience}, vol. 171,
  no.~1, pp. 196--204, 2010.

\bibitem{sakellari1997hypervent}
V.~Sakellari, A.~Bronstein, S.~Corna, C.~Hammon, S.~Jones, and C.~Wolsley,
  ``The effects of hyperventilation on postural control mechanisms,''
  \emph{Brain}, vol. 120, no.~9, pp. 1659--1673, 1997.

\bibitem{sorond2010cerebrovascular}
F.~Sorond, A.~Galica, J.~Serrador, D.~Kiely, I.~Iloputaife, L.~Cupples, and
  L.~A. Lipsitz, ``Cerebrovascular hemodynamics, gait, and falls in an elderly
  population: {MOBILIZE Boston} study,'' \emph{Neurology}, vol.~74, no.~20, pp.
  1627--1633, 2010.

\bibitem{gupta2007oh}
V.~Gupta and L.~A. Lipsitz, ``Orthostatic hypotension in the elderly: diagnosis
  and treatment,'' \emph{The American Journal of Medicine}, vol. 120, no.~10,
  pp. 841--847, 2007.

\bibitem{serrador2000parabolic}
J.~Serrador, J.~Shoemaker, T.~Brown, M.~Kassam, R.~Bondar, and T.~Schlegel,
  ``Cerebral vasoconstriction precedes orthostatic intolerance after parabolic
  flight,'' \emph{Brain Research Bulletin}, vol.~53, no.~1, pp. 113--120, 2000.

\bibitem{novak1998hypocapnia}
V.~Novak, J.~M. Spies, P.~Novak, B.~R. McPhee, T.~A. Rummans, and P.~A. Low,
  ``Hypocapnia and cerebral hypoperfusion in orthostatic intolerance,''
  \emph{Stroke}, vol.~29, no.~9, pp. 1876--1881, 1998.

\bibitem{rowell1993}
L.~B. Rowell, \emph{Human Cardiovascular Control}.\hskip 1em plus 0.5em minus
  0.4em\relax Oxford University Press, 1993.

\bibitem{gribbin1971baroreflex}
B.~Gribbin, T.~G. Pickering, P.~Sleight, and R.~Peto, ``Effect of age and high
  blood pressure on baroreflex sensitivity in man,'' \emph{Circulation
  Research}, vol.~29, no.~4, pp. 424--431, 1971.

\bibitem{kornet2005baroreflex}
L.~Kornet, A.~P. Hoeks, B.~J. Janssen, A.~J. Houben, P.~W. De~Leeuw, and R.~S.
  Reneman, ``Neural activity of the cardiac baroreflex decreases with age in
  normotensive and hypertensive subjects,'' \emph{Journal of Hypertension},
  vol.~23, no.~4, pp. 815--823, 2005.

\bibitem{low2015ohmechanisms}
P.~A. Low and V.~A. Tomalia, ``Orthostatic hypotension: mechanisms, causes,
  management,'' \emph{Journal of Clinical Neurology}, vol.~11, no.~3, pp.
  220--226, 2015.

\bibitem{finucane2017oh}
C.~Finucane and R.~A. Kenny, ``Falls risk, orthostatic hypotension, and optimum
  blood pressure management: is it all in our heads?'' \emph{American Journal
  of Hypertension}, vol.~30, no.~2, pp. 115--117, 2017.

\bibitem{arning2015camera}
K.~Arning and M.~Ziefle, ``{``Get that camera out of my house!''} {C}onjoint
  measurement of preferences for video-based healthcare monitoring systems in
  private and public places,'' in \emph{International Conference on Smart Homes
  and Health Telematics}, 2015, pp. 152--164.

\bibitem{sakellari1997hyperventilation2}
V.~Sakellari and A.~M. Bronstein, ``Hyperventilation effect on postural sway,''
  \emph{Archives of Physical Medicine and Rehabilitation}, vol.~78, no.~7, pp.
  730--736, 1997.

\end{thebibliography}

\newpage

\end{document}